I.V. GURYEV[1]
O.V. SHULIKA[1]
I.A. SUKHOIVANOV[1,2,✉]
O.V. MASHOSHINA[1]

# Improvement of characterization accuracy of the nonlinear photonic crystals using finite elements-iterative method

[1] Lab. "Photonics", National University of Radio Electronics, Kharkiv, 61166, Ukraine
[2] FIMEE, University of Guanajuato, Salamanca, 36790, Mexico

**ABSTRACT** We investigate nonlinear one- and two-dimensional photonic crystals by applying a finite element-iterative method. Numerical results show the essential influence of nonlinear elements embedded into a quarter-wave stack and the sharp photonic crystal waveguide bend on the spectral characteristics of these structures. We compare our results with those obtained in [21] from the discrete equation method for the case of a sharp waveguide bend. The comparison shows that neglecting the nonuniform field distribution inside the embedded nonlinear elements leads to overestimation of the waveguide bend transmissivity.

## 1    Introduction

Currently one of the most relevant problems in the field of information processing systems is connected with the fundamental limitations of semiconductor logical elements and with power supply requirements [1]. The optimal solution for this problem can be found by looking for new information handling principles. One of the alternative ways is the creation of all-optical devices. Periodic structures with photonic band gaps (PBG) – photonic crystals (PhC) [2], can be used as a medium for the construction of such devices. Due to their strict periodicity, the effect of strong light localization in the defect region of the structure appears. The theoretical and the experimental investigations demonstrate that the PhCs can be applied for lateral mode control in VCSELs [3], for creation of high-Q nanocavities [4], optical waveguides and sharp bends [5], WDM-devices [6, 7], splitters and combiners [8]. Thus, there is the possibility of building a fully functional optical processor on a single PhC structure [9].

Another attractive application of PhCs is the creation of ultracompact photonic integrated circuits. It would provide seamless, all-optical signal processing capabilities. This concept rests on the ability of certain PhCs to eliminate the propagation of light over extended frequency ranges irrespective of the propagation direction [10].

The aim of the work is to analyze the optical properties of nonlinear 1D and 2D periodic structures. For this reason, the spectral characteristics of 1D and 2D PhCs are investigated using the iterative method together with finite elements method.

## 2    Method of investigation

The effect of the nonlinear (NL) variation of the material permittivity can be produced by several factors. The most essential of them are

1. non linear molecular polarization of the material (for instance, in KDP, ADP);
2. carrier induced nonlinearity;
3. thermal variation of the permittivity.

Here we consider the optical Kerr-effect only. In this case the refractive index has a quadratic dependence upon intensity [9, 11]. The law of permittivity variation in its general form is as follows:

$$\varepsilon(E) = \varepsilon_l + \varepsilon_{nl}(E) . \tag{1}$$

Here $\varepsilon_l$ is the linear part of the relative permittivity, and $\varepsilon_{nl}(E)$ is the nonlinear part of permittivity. It is defined as $\varepsilon_{nl}(E) = a_{nl}|E|^2$, where $a_{nl}$ is the nonlinear coefficient.

The NL Helmholtz equation for this case is obtained by the substitution of (1) into the stationary Helmholtz equation. After this, it takes the form of:

$$\Delta\boldsymbol{E} + \left(\frac{\omega}{c}\right)^2 \varepsilon_l \boldsymbol{E} = -\left(\frac{\omega}{c}\right)^2 \varepsilon_{nl}(E)\boldsymbol{E} . \tag{2}$$

For the solution of (2), we use the iterative method [12] with finite elements method (FEM) [13]. FEM allows treating of structures of arbitrary complexity. This advantage is important when dealing with non-regular structures such as PhCs with embedded defects.

The method consists of successive refinement of the solution using the electrical field distribution in a structure. By solving the NL equation in such a way, we achieve a full correspondence between the field distribution and the RI spatial distribution in NL elements.

✉ Fax: +52-464-647-2400, E-mail: i.sukhoivanov@ieee.org

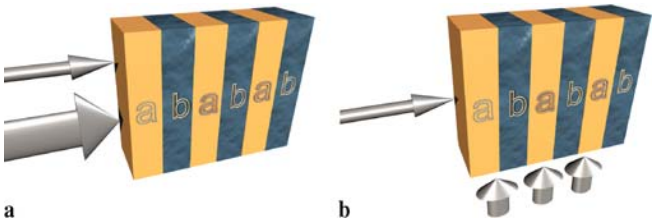

**FIGURE 1** The scheme of 1D NL PhC: "a"-linear layer; "b"-NL layer (**a**) co-directional propagation of pump and probe signals; (**b**) orthogonal propagation of pump and probe signals

The procedure itself consists of four steps.

(i) Initializing the whole device by setting the electric field intensity equal to zero in the investigated area.

(ii) Finding an FEM-solution of (2).

(iii) Computation of the spatial distribution of the permittivity.

(iv) Return to step (ii).

The end of the iteration process is determined by the difference between the varying parameters in the previous and the next iteration. When the difference is less then the tolerance value, the iteration process terminates. The tolerance value can be chosen depending on the accuracy required for the problem.

This technique allows treating with larger range of intensities as against analytical or semi-analytical methods. Accuracy depends only on the computation time. However, it can be difficult to achieve convergence at significantly high induced nonlinearity.

All-optical device with NL elements have typically two different operating regimes, linear and nonlinear. The physics of the device both quantitatively and qualitatively differs in these two regimes.

## 3    One-dimensional NL PhC

One-dimensional (1D) PhC represents a layered medium whose optical properties are determined by (i) layer thickness, (ii) RI contrast, and (iii) the reaction of layers on the optical field in a structure. In the case of NL 1D PhC, step (iii) plays the main role and is determined by the type of nonlinearity. The 1D PhC under investigation represents the quarter-wavelength Bragg stack formed by three pairs of layers (Fig. 1) composed of two types of materials. The first one possesses only linear optical properties while the second one demonstrates an optical nonlinearity. We use many NL layers as we expect in this case the largest reaction of the structure on intensity variation. The widths and linear permittivity of layers are $d_a = 0.256\,\mu m$, $d_b = 0.126\,\mu m$, $\varepsilon_l^a = 2.13$, $\varepsilon_l^b = 9.36$ respectively. Such structure is designed to have the Bragg wavelength of 1.55 μm in linear regime.

We analyze (i) co-directional and (ii) orthogonal configurations of the pump and probe signals, as depicted in Fig. 1. Probe signal incidents perpendicularly to the layers in both configurations. In case (i) the pump signal is injected perpendicularly to the layers (Fig. 1a), i.e. the pump and probe are co-propagating. In case (ii) the propagation direction of the pump signal is paralleling to the interfaces (Fig. 1b), i.e. the pump and probe are orthogonal. The pump and probe have the same wavelength in both configurations.

Reflection spectra corresponding to the described configurations are depicted in Fig. 2. Parameter $I$ is defined as $I = a_{nl} E_0^2$ with $E_0$ being the amplitude of the electric field at the input of PhC. This normalization allows to use these results for a large class of devices with scaled intensity.

The FWHM of the reflectivity widens by 0.4 μm in the NL regime ($I = 0.5$) and co-directional propagation (Fig. 2a). The widening also depends on the layers number. The pump wavelength is in the photonic band-gap. Therefore, the penetration depth of the pump signal is small due to the Bragg reflection. Hence the NL layers towards the end of stack are less influenced. Due to this fact chirping of the structure will take place, and the total structure reflection spectrum will widen even more than in the case of a low number of layers.

The Bragg wavelength of the layered structure can be written as [14]:

$$\lambda_B = 2\,(n_a\,d_a + n_b d_b)\ , \qquad\qquad (3)$$

where the $d_a$ and $d_b$ are the widths of linear and nonlinear layers respectively, and $n_a$ and $n_b$ are their total RIs. From (3) it follows that an increase of the RIs contrast leads to an increase of the Bragg wavelength. The intensity growth of the pump signal in both configurations results in a growth of the average RI contrast. Therefore, the maximum of the reflection spectrum is red-shifted.

There are wavelengths where the reflectivity is almost zero ($\lambda = 2.57\,\mu m$). When increasing the intensity, the spectrum shape changes so that the reflectivity at these wavelengths is no longer equal to zero. In the co-directional configuration increasing the reflectivity leads to a decrease of the penetration depth. This results in saturation of the growth of reflectivity at these wavelengths.

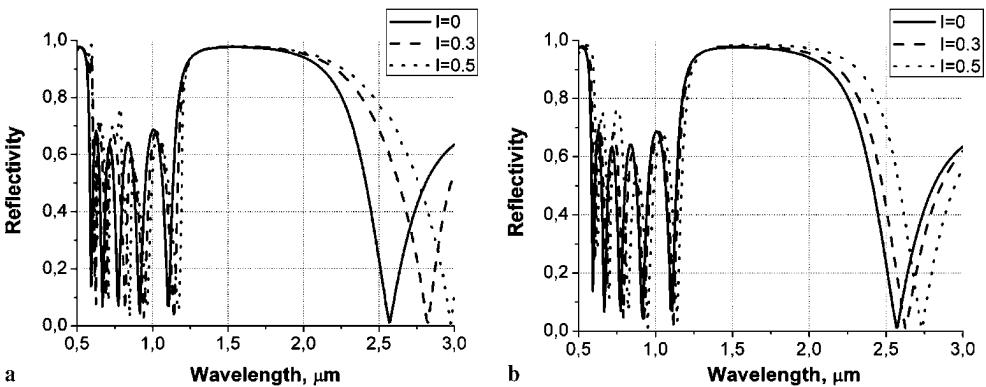

**FIGURE 2** Reflection spectra in linear and NL regimes: (**a**) co-directional propagation of pump and probe signals; (**b**) orthogonal propagation of pump and probe signals

However, the situation is the reverse at the wavelengths that correspond in linear regime to one of the local reflectivity maxima excluding the main maximum (for example $\lambda = 1.175\,\mu m$). When increasing the intensity the spectrum shape changes so that the reflectivity at the selected wavelengths decreases. The penetration depth therefore increases.

The NL effect does not influence reflectivity at the Bragg wavelengths because they are always in PBG, both in linear and NL regimes (see Fig. 3).

The shape of the curves in Fig. 3 depends only on the shape of the reflectivity spectrum for the selected wavelength and around it. When one needs a large variation of reflectivity, it is most useful to take wavelengths at narrow transmission or reflection peaks.

## 4    PhC waveguide with nonlinear insertion

The typical property of the 2D PhC is the existence of PBG in the frequency domain for the radiation of specific polarization propagating in a specific plane [15]. Such structures are usually formed by the periodic pattern of holes in semiconductors (dielectric) or the periodic pattern of semiconductor (dielectric) rods in air or any other low-RI material. It is possible to create localization conditions for radiation at a certain wavelength range by forming a defect or rows of defects inside the periodic structure. Moreover, such defects allows realization of low-loss sharp waveguide bends [16], splitters and add/drop multiplexers [17]. Embedding the nonlinear elements into 2D PhC-waveguides, allows creating of devices demonstrating optical bistability [18], unidirectionality [19] as well as all-optical logical items [20]. From the point of view of understanding the nonlinear effect, it is convenient to use simple structures in which results can be explained intuitively. Therefore, we investigated simple a 2D W1 waveguide channel with embedded nonlinear insertions and a sharp waveguide bend based on the same type of PhC .

Depending on the specific problem, the initial parameters of insertions can be the same as for background PhC's

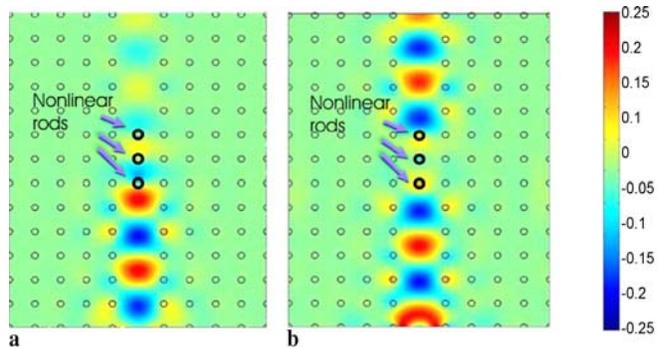

**FIGURE 4**    Computed electrical field distribution: (**a**) linear regime (**b**) nonlinear regime. The rods marked with *bold circles* exhibit nonlinearities

and differs from them as well. We have chosen parameters so as to provide partial overlapping of the PBGs of background PhC and PhC of insertions. It provides the initial state of device to be "closed" in the linear regime, i.e. no radiation passes through the filter. In the NL regime the filter will be "opened", i.e. we will observe a strong increase of the transmission.

The following parameters are used in the numerical simulations. Background PhC: square lattice of pillars, lattice constant $a = 0.57\,\mu m$; pillar radius $r = 0.075\,\mu m$, RI of pillars $n = 3.5$. NL insertions represent three identical rods with radii $r_{nl} = 0.12\,\mu m$ possessing optical nonlinearity.

The field distribution in the structure was found via the solution of (2) using the finite elements-iterative method as was described in Sect. 2. The mesh for this case consists of about $3 \times 10^{4}$ finite elements. The computation time on an AMD 2200+ processor is about 30 s for the linear regime and 2–20 min for the NL regime depending on the light intensity. Here and in the next section the number of iterations needed to achieve convergence varies from 5 to 40.

Results of computation of the electric field distribution in linear ($I = 0$) and nonlinear ($I = 4$) regimes at $\lambda = 1.375\,\mu m$ are shown in Fig. 4. Here the parameter $I$ has the same meaning as in Sect. 3.

As is seen from Fig. 4, increasing the incident intensity results in an essential increase of transmission. This is due to the spectral shift of the filter PBG.

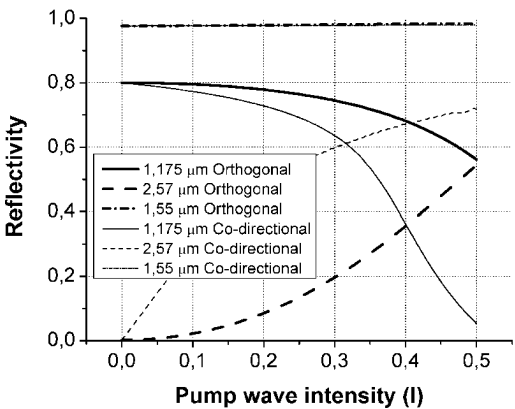

**FIGURE 3**    The dependence of the reflectivity on light intensity at different wavelengths

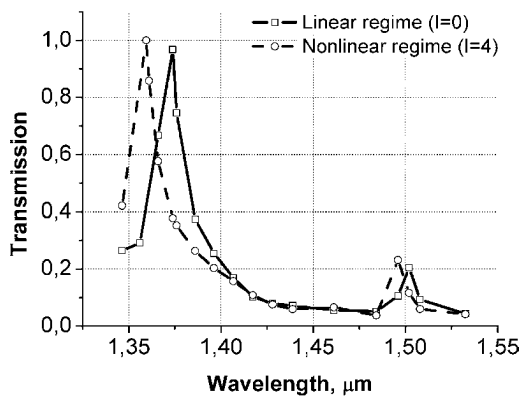

**FIGURE 5**    The transmission spectrum of the straight waveguide with embedded nonlinear insertions in linear and nonlinear regimes

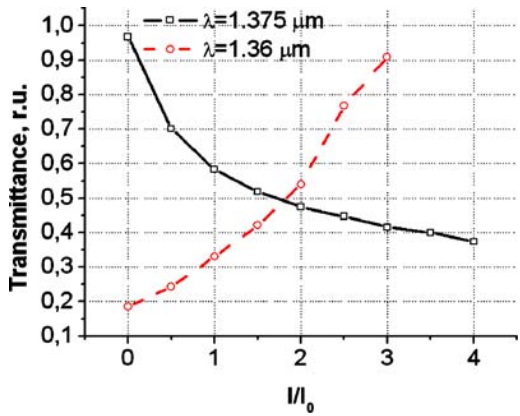

**FIGURE 6**   Dependence of transmission on the light intensity

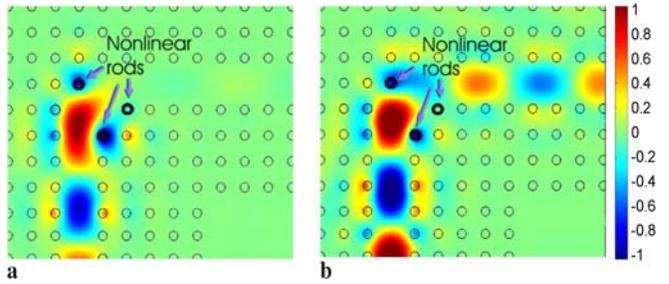

**FIGURE 7**   Electrical field distribution in a nonlinear bend (bold lines indicate nonlinear elements): (**a**) linear regime; (**b**) nonlinear regime

Transmission spectra of the waveguide in linear and nonlinear regimes are shown in Fig. 5. The computation was made in the spectral range which includes the PBGs of both background PhC and filter PhC. The spectral range is chosen to satisfy light localization conditions inside the waveguide.

We found two transmission peaks within the selected spectral range in both the linear and nonlinear regimes. Transmission of the short-wave peak is 99%. Transmission maxima are blue-shifted in NL regime.

At the wavelength corresponding to the transmission maximum in the linear regime ($\lambda = 1.375\,\mu m$), the transmission in the NL regime is 35%. However, at the wavelength corresponding to the transmission maximum in the NL regime ($\lambda = 1.36\,\mu m$), the transmission in the linear regime achieves only 20%. Thus, these wavelengths have an almost equal ratio between transmissions in the linear and nonlinear regimes. From a practical point of view, it means in principal a possibility for the creation of two devices with opposite functions (for instance "AND" and "NOT-AND") based on the same structure.

The transmission of nonlinear insertions as a function of the incident intensity is shown in Fig. 6. Wavelengths correspond to maximum transmission in linear and NL regimes. One can see that the change of transmission on both wavelengths is sub-linear. It can cause some problems if we want to work with a smooth variation of the light intensity. The value of the incident intensity can be reduced via additional structural optimization.

## 5   Sharp PhC-bend with nonlinear elements

The background PhC has square lattice with the lattice constant $a = 0.57\,\mu m$; pillar radius $r = 0.102\,\mu m$, RI of pillars $n = 3.4$, total RI of nonlinear pillars $n = 2.64$.

The computed electrical field in a sharp PhC-bend is shown in Fig. 7. The number of elements is about $3.5 \times 10^4$. The computation time on an AMD 2200+ processor is about 50 s for the linear regime and 3–30 min for the nonlinear regime, depending on the light intensity.

The computations are for $\lambda = 1.63\,\mu m$ because of highest transmission contrast at this wavelength (see the spectrum computation below). As we can see in Fig. 7, the sharp PhC-bend has a strong dependence of transmission on the intensity at certain wavelengths as in the case of the straight PhC-waveguide.

The transmission spectra in linear and nonlinear regimes were obtained by a gradual change of the wavelengths in the spectral range under consideration. In the linear regime, we obtained transmission spectra identical to those presented in [21]. The transmission spectrum of the sharp PhC-bend in [21] has the same shape in both linear and NL regimes and is only red-shifted in the NL regime. This is due to uniform field distribution inside the pillars as supposed in [21]. However, the field distribution inside the pillars can be appreciably non-uniform because the pillar section is comparable to the operating wavelength. Field non-uniformity is enhanced in the NL regime. Figure 8a demonstrates the contour plot of the electrical field distribution in a NL pillar and Fig. 8b presents the field magnitude in a diametral section A of Fig. 8a. The field magnitude is normalized as $E' = E\sqrt{a_{nl}}$. It was made to avoid associations with any material.

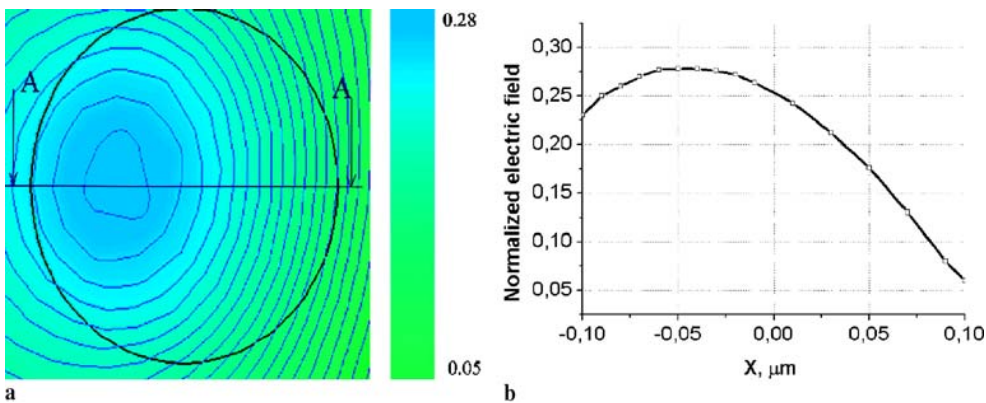

**FIGURE 8**   Electrical field distribution inside the NL pillar in the NL regime. (**a**) contour plot, (**b**) in diametral cross-section A ($x = 0$ corresponds to element center)

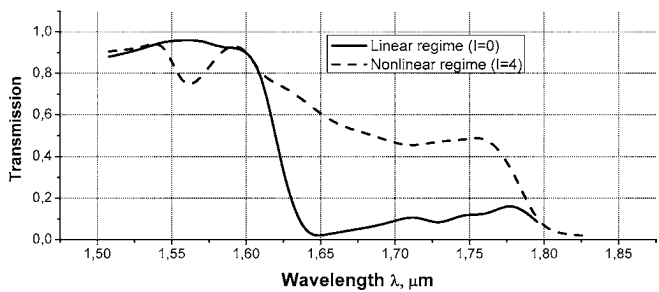

We found field non-uniformity in a change of the transmission spectrum in the NL regime (Fig. 9). At the same time there is no step of the transmission in the range 1.6–1.65 μm. Instead, we observed an increase of transmission to values of 0.5–0.7 around the wavelengths 1.6–1.76 μm. Violation of the localization condition takes place at longer wavelengths. This leads to the flow of radiation out of the structure and, hence, device failure.

Thus, our computations have shown that field distribution cannot be neglected inside the NL rods if optical nonlinearities are to be considered.

Our analysis gives an introduction to the application of sharp NL PhC-bends in photonic integrated circuits. Just like a straight PhC-waveguide with NL embeddings sharp NL PhC-bend can be applied as the logical cell. However, the low contrast of transmission in the linear and NL regimes make it less preferable.

## 6        Conclusion

We investigated 1D and 2D PBG-structures with optical Kerr nonlinearity. Electrical field distributions, and spectral characteristics in linear and nonlinear regimes were computed. We found in the NL regime a shift of the transmission peak at 20 nm in the 2D case as well as widening of the main reflection peak at 700 nm in the 1D case. We found that the switching efficiency of the sharp PhC-bend is almost equal to the switching efficiency of the straight PhC-waveguide. However, in contrast to the straight waveguide PhC-bend it does not provide the possibility for creation of logical elements with opposite functions. We have shown that the field distribution inside the nonlinear elements can not be neglected in modeling and simulation of nonlinear PhCs.